# Domain knowledge aided machine learning method for properties prediction of soft magnetic metallic glasses

Xin LI[1,2], Guang-cun SHAN*[1,2], Hong-bin ZHAO[3], Chan Hung SHEK[2,*]

1．School of Instrumentation Science and Opto-electronics Engineering, Beihang University, Beijing 100191, China

2. Department of Materials Science and Engineering, City University of Hong Kong, Kowloon Tong, Hong Kong SAR, China

3. State Key Laboratory of Advanced Materials for Smart Sensing, General Research Institute for Nonferrous Metals, Beijing 100088, China

*Corresponding author: Guang-cun SHAN, E-mail: gcshan@buaa.edu.cn; Chan Hung SHEK, E-mail: apchshek@cityu.edu.hk

# Abstract

A machine learning (ML) method aided by domain knowledge was proposed to predict saturated magnetization ($B_s$) and critical diameter ($D_{max}$) of soft magnetic metallic glass (MGs). Two datasets were established based on published experimental works about soft magnetic MGs. A general feature space was proposed and proved to be adaptive for ML model training for different prediction tasks. It was found that the predictive performance of ML models was better than traditional physical knowledge-based estimation methods. In addition, domain knowledge aided feature selection can greatly reduce the number of features without significantly reducing the prediction accuracy. Finally, binary classification of the critical size of soft magnetic metallic glass was studied.

# 1. Introduction

Soft magnetic metallic glasses (MGs) have attracted considerable attention due to their unique combination of excellent mechanical and magnetic properties [1]. Due to the amorphous structure, soft magnetic MGs usually do not have magnetic anisotropy and grain boundaries, which make them exhibit excellent soft magnetic properties, i.e., high permeability and saturated magnetization ($B_s$), and low coercivity. For functional engineering application of soft magnetic MGs, the main goal is to improve $B_s$. $B_s$ is an intrinsic property of magnetic materials, which refers to the maximum flux density can be achieved. In general, MGs with high $B_s$ value have a high content of ferromagnetic elements, for example, 1.75 T for $Fe_{86}B_7C_7$ ribbon [2], 1.82 T for $(Fe_{0.8}Co_{0.2})_{87}B_7Si_3P_3$ ribbon [3], 1.9 T for $(Fe_{0.8}Co_{0.2})_{85}B_{14}Si_1$ ribbon [4], 1.51 T for $Fe_{76}Si_9B_{10}P_5$ rod with a critical diameter ($D_{max}$) of 2.5 mm [5], 1.61 T for $Fe_{75.3}C_7Si_{3.3}B_5P_{8.7}Cu_{0.7}$ rod with a $D_{max}$ of 1.5 mm [6]. $D_{max}$ is the diameter of the largest amorphous rod that an alloying composition can form. It is an experimental parameter that could be easily obtained and is often used to quantify glass forming ability (GFA) of MGs. Though ~ 5000 MGs compositions have been developed [7], most of them were in a ribbon form and showed too limited GFA to form bulk MGs with large geometric size. Therefore, understanding $B_s$ and $D_{max}$ of soft magnetic MGs is an important research topic, and prediction of the two properties is of great significance for the design of high performance soft magnetic MGs.

In the past few years, machine learning (ML) has been applied to material science to predict material properties or behaviors [8–10]. GFA and magnetic property of alloys were also have been studied by ML methods, which can be divided into classification and regression. For classification analysis, ML models based on backpropagation neural network [11], support vector machine [12], general and transferable deep learning framework [13], etc., were trained to identify MGs and non-MGs classes. For regression analysis, ML models were trained to predict the specific $D_{max}$ value of MGs[14,15], and $B_s$ value of $MG_s$ [16] and nanocrystalline alloys [17].

In this work, a ML method aided by domain knowledge was proposed to predict $B_s$ and $D_{max}$ of soft magnetic MGs based on alloying compositions. Two datasets were established based on previous work. ML models were trained based on five ML algorithms, and their predictive performance was compared. A general feature space was proposed for the prediction of the two properties, and feature selection based on domain knowledge was conducted. By comparing the predictive performance of ML models before and after feature selection, the effect of domain knowledge in the ML method was highlighted.

## 2. Methodology

### 2.1. Dataset description

The two datasets were deduced from previous work. A data entry in the datasets contains the information about the elemental components and the experimental values of target properties, i.e., $B_s$ with a unit of T and $D_{max}$ with a unit of mm. The saturated magnetization dataset (hereinafter referred to as BS Dataset) contains 639 MGs compositions, most of them were taken from a previous related work [18], and the rest were collected from other published literatures [19–23]. Since the average magnetic moment of iron is larger than cobalt and nickel, and the cost of production of the former is lower than the latter, Fe-based MGs become the mainstream in soft magnetic MGs. The critical casting diameter dataset (hereinafter referred to as DMAX Dataset) contains 519 Fe-based alloying compositions, which were selected from the two previous related works [14,24]. The literature [24] provided a Fe-based bulk MGs dataset containing 480 alloying compositions with experimental $D_{max}$ values. Another literature [14] provided a bulk MGs dataset containing 7950 alloying compositions with experimental $D_{max}$ values, out of which 139 are Fe-based bulk MGs. The $D_{max}$ values of the same alloying composition reported in different works may be different, in this case, the average value was adopted.

Figures 1(a) and 1(b) show the value distribution of $B_s$ and $D_{max}$, respectively. The value of $B_s$ in the dataset ranges from 0.05 T to 1.92 T with a median of 1.5 T and an average of 1.19 T, which shows the excellent magnetic property of MGs. The value of $D_{max}$ in the dataset ranges from 0.06 mm to 18.00 mm with a median of 2.50 mm and an average of 3.04 mm. The $D_{max}$ data is sparsely distributed in the interval greater than 7 mm, which indicates that improving GFA of MGs is still challenging. Figures 1(c) and 1(d) show the distribution of the chemical elements in the two datasets. There are 33 and 29 kinds of chemical elements in BS Dataset and DMAX Dataset, respectively, and ~ 95 % of the alloying compositions contain Fe and B. As mentioned before, Fe is the main ferromagnetic element used in soft magnetic MGs. Proper addition of boron could enhance GFA of Fe-based MGs without excessively negative effect on the magnetic property [25]. Figures 1(c) and 1(d) also indicate that there are a lot of chemical elements could be used for the preparation of soft magnetic MGs, therefore, an effective properties prediction method is of great significance for the development of new soft magnetic MGs.

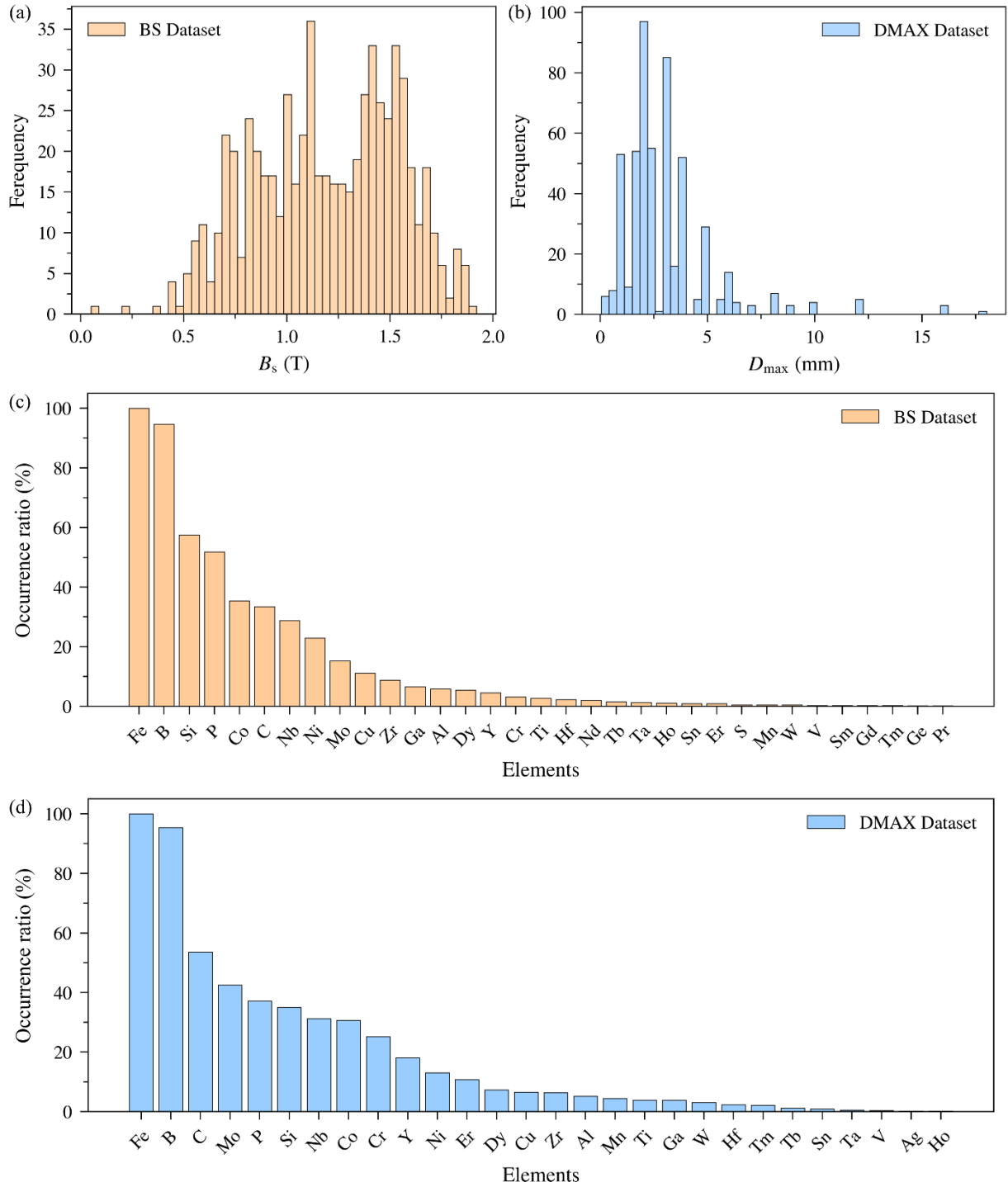

Figure 1. Distribution of $B_s$ (a), $D_{max}$ (b), and chemical elements in BS Dataset (c) and DMAX Dataset (d).

In general, soft magnetic alloys can be easily magnetized and demagnetized, which is usually identified by whether the values of their intrinsic coercivity ($H_c$) are less than 1 kA/m.

A scatter plot of the values of $B_s$ versus $H_c$ of the alloying compositions in BS Dataset is shown in Figure 2(a), which indicates that they exhibit typical soft magnetic property. Most of the alloying compositions have a $H_c$ value even lower than 50 A/m. Considering both datasets, there are 130 alloying compositions with both $B_s$ and $D_{max}$ values. As shown in Figure 2(b), $B_s$ and $D_{max}$ values of these alloying compositions show an obviously negative correlation. Their Pearson correlation coefficient was calculated to be -0.10, which confirms the negative correlation. The calculation method of Pearson correlation coefficient could be found in the previous work [18]. Therefore, in general, magnetic property and GFA of soft magnetic MGs are incompatible with each other. Understanding and modeling the two properties could help to design high performance soft magnetic MGs.

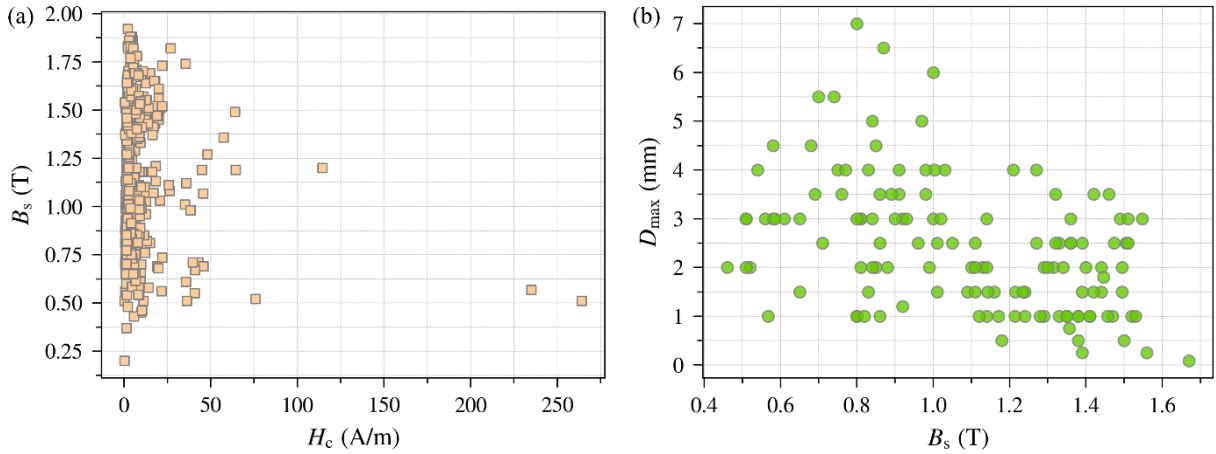

Figure 2. Scatter plots of $B_s$ versus $H_c$ (a) and $D_{max}$ versus $B_s$ (b).

## 2.2 General feature space

Feature space is a combination of some measurable parameters used to describe the input, which is alloying compositions in this work, of ML models. For designing new functional alloys with desired properties, it is more practical not to involve any parameters about synthesis process or experimental measurements. Therefore, the features used in this work to describe soft magnetic MGs compositions were all only based on chemical components and elemental properties. A general feature space was proposed to conduct ML models training and prediction of both $B_s$ and $D_{max}$. Apart from the feature candidates used in the previous work [18], including elements and atomic percent ($c$), theoretical density ($\rho$), mean metallic atomic radius [26] ($R_m$), atomic size difference ($\delta_R$), theoretical molar volume [27] ($V_m$), melting temperature calculated by the rule of mixtures ($T_m$), Pauling electronegativity [26] ($\chi$), valence electron concentration [28] (*VEC*) and mixing entropy [29] ($\Delta S_{mix}$), more feature candidates were involved in this work, including mixing enthalpy [30] ($\Delta H_{mix}$), mixing Gibbs free energy [14] ($\Delta G_{mix}$), Pauling electronegativity difference ($\delta_\chi$), relative Pauling electronegativity ($R\chi$), valence electron concentration difference ($\delta_{VEC}$), relative valence electron concentration (*RVEC*), work function [31] ($W$) and work function difference ($\delta_W$), as shown in Table 1. All the feature candidates in the general feature space can be divided into 4 classes related to chemical components, features related to atomic structures, features related to thermodynamics and features related to electronic properties. The feature candidates related to atomic structures, thermodynamics and electronic properties mainly describe the atomic percent weighted average and mismatch of the elemental properties in alloying compositions by

$$\bar{r} = \sum_{i=1}^{n} c_i p_i \qquad (1)$$

$$\delta = \sqrt{\sum_{i=1}^{n} c_i \left(1 - \frac{p_i}{\bar{p}}\right)^2} \qquad (2)$$

where $n$ is the total number of chemical elements in an alloying composition, and $c_i$ and $p_i$ is the atomic fraction and elemental properties of the $i$-th element, respectively. In addition, $\Delta S_{mix}$, $\Delta H_{mix}$, $\Delta G_{mix}$, $R\chi$ and $RVEC$ are calculated by

$$\Delta S_{mix} = -R \sum_{i=1}^{n} c_i \ln c_i \qquad (3)$$

$$\Delta H_{mix} = \sum_{i=1}^{n} 4 c_i c_j \Delta H_{AB}^{mix}, i \neq j \qquad (4)$$

$$\Delta G_{mix} = \Delta H_{mix} - T_m \Delta S_{mix} \qquad (5)$$

$$R\chi = \sum_{i=1}^{n} c_i \chi_i - c_{Fe}\chi_{Fe} - c_{Co}\chi_{Co} - c_{Ni}\chi_{Ni} - c_{Gd}\chi_{Gd} \qquad (6)$$

$$RVEC = \sum_{i=1}^{n} c_i VEC_i - c_{Fe} VEC_{Fe} - c_{Co} VEC_{Co} - c_{Ni} VEC_{Ni} - c_{Gd} VEC_{Gd} \qquad (7)$$

where $c_i$, $\chi_i$ and $VEC_i$ are the atomic fraction, electronegativity, valence electron concentration of the $i$-th element, and $c_j$ is the atomic fraction of the $j$-th element. R is the gas constant, and $\Delta H_{AB}^{mix}$ is the mixing enthalpy of the liquid binary alloy containing $i$-th and $j$-th elements. $\chi_{Fe} = 1.83$, $\chi_{Co} = 1.88$, $\chi_{Ni} = 1.91$, $\chi_{Gd} = 1.2$, $VEC_{Fe} = 8$, $VEC_{Co} = 9$, $VEC_{Ni} = 10$ and $VEC_{Gd} = 3$.

Table 1. The general feature space used for ML models training and prediction.

| Feature class | Description |
| --- | --- |
| Chemical components | Atomic percent of chemical elements ($c$) |
| Related to atomic structures | Theoretical density ($\rho$) |
| | Mean metallic atomic radius ($R_m$) |
| | Atomic size difference ($\delta_R$) |
| | Theoretical molar volume ($V_m$) |
| Related to thermodynamics | Mixing entropy ($\Delta S_{mix}$) |
| | Mixing enthalpy ($\Delta H_{mix}$) |
| | Mixing Gibbs free energy ($\Delta G_{mix}$) |
| | Melting temperature calculated by the rule of mixtures ($T_m$) |
| Related to electronic properties | Pauling electronegativity ($\chi$) |
| | Pauling electronegativity difference ($\delta_\chi$) |
| | Relative Pauling electronegativity ($R\chi$) |
| | Valence electron concentration ($VEC$) |
| | Valence electron concentration difference ($\delta_{VEC}$) |
| | Relative valence electron concentration ($RVEC$) |
| | Work function ($W$) |
| | Work function difference ($\delta_W$) |

## 2.3 Knowledge-based feature selection

The general feature space contained as much information as possible to be applicable to different prediction tasks. However, too many features might hide the physical mechanism dominating the target properties, and redundant information also has a negative impact on the predictive performance of ML models. Therefore, feature selection was conducted based on a voting strategy and domain knowledge. First, several prevalent feature selection methods from machine learning community were used to reduce the number of features for $B_s$ and $D_{max}$ prediction without significant loss of accuracy. Then, knowledge-based feature was added to improve the predictive performance of ML models. In the first feature selection process, four feature selection algorithms were used, namely univariate feature selection [32], feature importance given by light gradient boosting machine (LightGBM) [33] and recursive feature elimination (RFE) [34], which were denoted as M1, M2 and M3, respectively. Univariate feature selection could figure out the features having the strongest relationship with the target variable by statistical tests. The statistical test used in this work was the analysis of variance with the F-test [32]. LightGBM is a decision tree-based algorithm, which could produce feature importance by traversing each node of the established trees with a criterion. RFE recursively removes unimportant features suggested by ML models and trains a new ML model using the remaining features in the feature list. The ML algorithm used for producing feature importance in RFE was also LightGBM. The first feature selection procedure is mainly based on the principle of statistical analysis, which is data sensitive. In addition, many researchers have studied magnetic property and GFA of MGs from a material science point of view, which can be used to guide feature selection. For example, according to the theories in magnetism, $B_s$ is proportional to the average magnetic moment of all the atoms in an alloy [35], which is described by

$$B_s = \frac{N_A \bar{\mu} \mu_B}{V_m} \qquad (8)$$

where $N_A$ Avogadro constant, $\bar{\mu}$ is the mean magnetic moment of the alloy, $\mu_B$ is Bohr magneton, and $V_m$ is the molar volume of the alloy. Furthermore, since the 1930s, some theoretical estimation methods for the mean magnetic moment of an alloy had been continuously developed [36]. Recently, based on free electron transfer theory, the literature [35] proposed an estimation method of $\bar{\mu}$ in Fe-based alloys using the chemical compositions. Inspired by that work, the estimated $\bar{\mu}$ is selected as a feature.

## 2.4 Machine learning algorithms

Kinds of ML algorithms have been successfully applied to solve material science problems,

but there is not a universal ML algorithm. Therefore, it is necessary to choose a suitable ML algorithm and implement hyperparameters optimization for good predictive performance in different problems. Five ML algorithms were selected to address the problem of $B_s$ and $D_{max}$ prediction in this work, namely support vector regression (SVR) [37], multilayer perceptron (MLP) [37], random forest (RF) [38], eXtreme gradient boosting (XGBoost) [39] and LightGBM. Their predictive performance was evaluated by 10-fold cross-validation [40], which is a prevalent method to objectively evaluate the generalization ability of ML models when the dataset size is limited. The evaluation metric used in cross-validation was root mean squared error (RMSE) calculated by

$$\mathrm{RMSE}(y_i, \hat{y}_i) = \sqrt{\frac{1}{n}\sum_{i=1}^{n}(y_i - \hat{y}_i)^2} \qquad (9)$$

where $y_i$ is the experimental value of $B_s$ or $D_{max}$, and $\hat{y}_i$ is the predicted values of $y_i$. The unit of RMSE is the same as the target variable, and the smaller the RMSE value, the better the predictive performance of the ML model. The final predictive performance was quantified by the average value of 10 RMSE scores produced by 10-fold cross-validation, and the ML algorithm with the lowest RMSE value was suggested as the most suitable algorithm for $B_s$ or $D_{max}$ prediction. In addition, to compare the prediction accuracy of the two target properties ($B_s$ and $D_{max}$) with different units, another metric, namely determination coefficient ($R^2$), was calculated by

$$R^2(y_i, \hat{y}_i) = 1 - \frac{\sum_{i=0}^{n}(y_i - \hat{y}_i)^2}{\sum_{i=0}^{n}(y_i - \bar{y})^2} \qquad (10)$$

where $y_i$ is the experimental values of $B_s$ or $D_{max}$, and $\bar{y}$ and $\hat{y}_i$ are the average and predicted values of $y_i$, respectively.

## 3. Results and discussions

### 3.1 ML models trained with general feature space

Based on the general feature space, the predictive performance of the 5 ML algorithms was tested via 10-fold cross-validation. The cross-validation results evaluated by RMSE and $R^2$ are shown in Figures 3(a) and 3(b), respectively. It was found that XGBoost ML models outperformed others with the lowest RMSE and highest $R^2$ scores for both $B_s$ and $D_{max}$ prediction. Figures 3(b) shows that the prediction accuracy for $D_{max}$ was much lower than that of $B_s$. The reason could be that the $D_{max}$ dataset is scattered, and the data density is too large for the range under 5mm, which indicates that GFA of most reported MGs is limited. In addition, experimental $D_{max}$ of MGs is measured by an injection casting method. The cooling rate of rods in copper mold casting is hard to keep uniform in different experiments. Therefore, the experimentally measured $D_{max}$ values could be fluctuated. The $R^2$ values of $B_s$ and $D_{max}$ prediction via XGBoost were ~ 0.93 and ~ 0.68, respectively, which were consistent with the recently reported results in MGs [16,18,24,41]. It should be noted that all the ML models were trained based on the same feature space, i.e., the general feature space mentioned above.

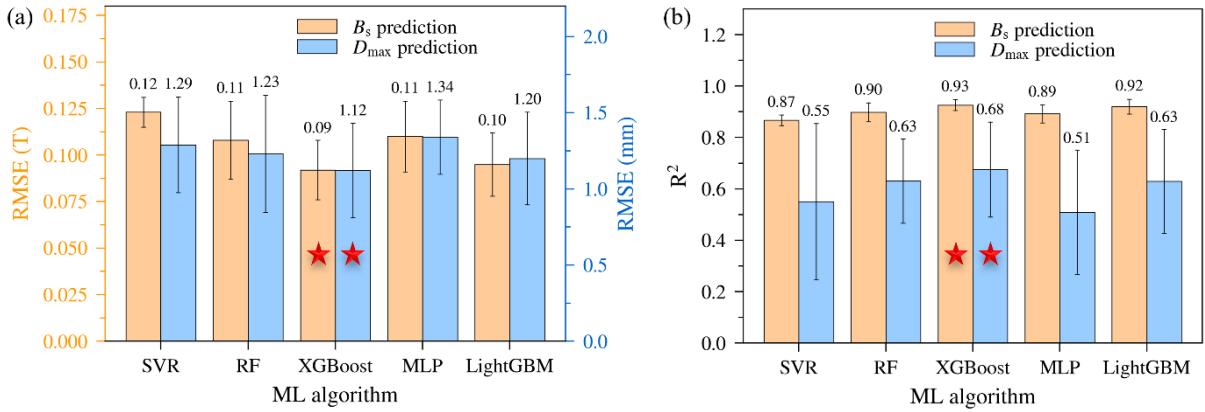

Figure 3. Cross-validation evaluated by RMSE (a) and $R^2$ (b) of different ML models.

## 3.2 ML models trained with selected feature space

Based on the general feature space, feature selection was conducted. In the first feature selection procedure, every feature selection strategy (M1, M2 and M3) was set to recommend 15 features out of the general feature space. Then final 10 features were selected based on the recommendations. The feature selection results for $B_s$ and $D_{max}$ prediction are listed in Table 2. It should be noted that the finally selected 10 features were not presented in any order of priority.

Table 2. Feature selection for the prediction of $B_s$ and $D_{max}$ in soft magnetic MGs via a voting strategy.

| Target | Method | Recommended features |
|---|---|---|
| $B_s$ | M1 | $\Delta G_{mix}$, $R\chi$, $RVEC$, $\Delta S_{mix}$, $c_{Fe}$, $\delta_{VEC}$, $c_{Nb}$, $\delta_R$, $c_{Ni}$, $\Delta H_{mix}$, $c_B$, $c_{Dy}$, $c_{Co}$, $T_m$, $\rho$ |
| | M2 | $T_m$, $\rho$, $\chi$, $\Delta G_{mix}$, $\Delta H_{mix}$, $VEC$, $\delta_{VEC}$, $W$, $R_m$, $\Delta S_{mix}$, $V_m$, $\delta_\chi$, $RVEC$, $R\chi$, $c_{Fe}$ |
| | M3 | $c_{Fe}$, $\rho$, $\delta_R$, $R\chi$, $\delta_{VEC}$, $RVEC$, $\Delta S_{mix}$, $T_m$, $\Delta H_{mix}$, $\Delta G_{mix}$, $VEC$, $\delta_\chi$, $R_m$, $V_m$, $\chi$, |
| | Final | $c_{Fe}$, $\rho$, $\delta_R$, $R\chi$, $\delta_{VEC}$, $RVEC$, $\Delta S_{mix}$, $T_m$, $\Delta H_{mix}$, $\Delta G_{mix}$ |
| $D_{max}$ | M1 | $\Delta G_{mix}$, $R\chi$, $RVEC$, $\delta_\chi$, $\Delta S_{mix}$, $c_{Cr}$, $T_m$, $VEC$, $c_{Mo}$, $\delta_R$, $c_{Tm}$, $\rho$, $c_C$, $\Delta H_{mix}$, $c_P$ |
| | M2 | $RVEC$, $T_m$, $\Delta H_{mix}$, $\Delta G_{mix}$, $\delta_\chi$, $\Delta S_{mix}$, $\chi$, $R_m$, $R\chi$, $VEC$, $\rho$, $\delta_{VEC}$, $\delta_R$, $W$, $c_{Fe}$ |
| | M3 | $\rho$, $R_m$, $\delta_R$, $\chi$, $\delta_\chi$, $R\chi$, $VEC$, $\delta_{VEC}$, $RVEC$, $V_m$, $\Delta S_{mix}$, $T_m$, $\Delta H_{mix}$, $\Delta G_{mix}$, $W$ |
| | Final | $\rho$, $\delta_R$, $\chi$, $\delta_\chi$, $R\chi$, $RVEC$, $\Delta S_{mix}$, $T_m$, $\Delta H_{mix}$, $\Delta G_{mix}$ |

Since XGBoost showed the best predictive performance in the general feature space, the impact of feature selection on its predictive performance was evaluated by cross validation using the same hyperparameters as before. Figure 4 shows the scatter plots of all the cross-validated predictions for $B_s$ and $D_{max}$. Comparing Figures 4(a) and 4(b), it can be found that the accuracy of $B_s$ prediction was slightly decreased after feature selection. Based on the final 10 features selected for $B_s$ prediction in Table 2, the estimated $\bar{\mu}$ [35] of each alloying compositions in BS Dataset was calculated and added as a feature. As shown in Figure 4(c), after adding this feature, the accuracy of $B_s$ prediction was improved a lot with $R^2$ increasing from 0.857 to 0.911, and RMSE decreasing from 0.127 T to 0.101 T. Therefore, the estimated $\bar{\mu}$ made a positive contribution to the predictive performance of XGBoost model. However, if $B_s$ was calculated directly by equation (8) based on the estimated $\bar{\mu}$, the accuracy would be much lower than XGBoost model. As shown in Figure 4(d), the $R^2$ and RMSE scores for this theoretical calculation method are only 0.547 and 0.230 T, respectively. The $\bar{\mu}$ estimation method for MGs was developed from free electron transfer theory, which is of great significance for understanding the magnetic property of MGs. However, to accurately predict magnetic property of MGs, more factors should be considered. According to the feature selection results, apart from features about electron transfer ($R\chi$, $\delta_{VEC}$ and $RVEC$), other features about structures ($c_{Fe}$, $\rho$ and $\delta_R$,) and thermodynamic properties ($\Delta S_{mix}$, $T_m$, $\Delta H_{mix}$ and $\Delta G_{mix}$) also had an impact on the prediction accuracy of $B_s$.

For $D_{max}$ prediction, the prediction accuracy just slightly decreased after feature selection, as shown in Figures 4(e) and (f). The $R^2$ slightly decreased from 0.675 to 0.669, and RMSE slightly increased from 1.121 mm to 1.119 mm. However, the number of features was greatly reduced from 44 to 10. A large reduction in the number of features is very helpful to improve the robustness of ML models and reduce the complexity of ML models. Though the prediction accuracy of $D_{max}$ via ML models was not as satisfactory as $B_s$, it had outperformed any traditional GFA descriptors based on experimental thermal dynamic parameters [41], for example, onset crystallization temperature ($T_x$), glass transition temperature ($T_g$), liquidus temperature ($T_l$) and supercooled liquid region ($\Delta T_x$). In addition, the ML-based GFA prediction method does not rely on any experimental parameters, which has more practical value in discovering new MGs.

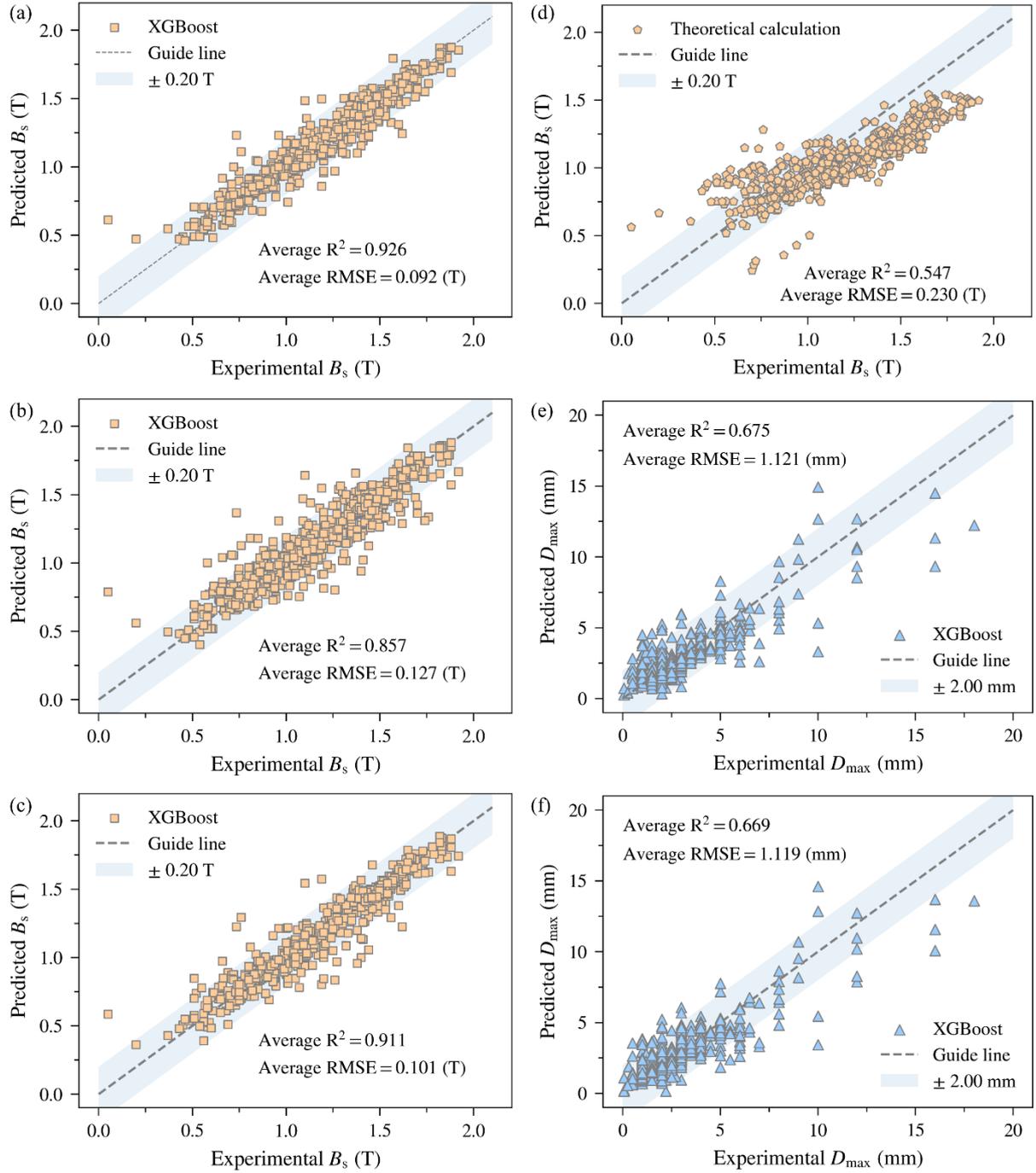

Figure 4. Cross-validated predictions of $B_s$ based on the general feature space (a), selected 10 features (b), selected 11 features with $\bar{\mu}$ (c) and theoretical calculation (d), and $D_{max}$ based on the general feature space (e) and selected 10 features (f).

More details in 10-fold cross validation are shown in Figure 5. The same 10-fold dataset splitting was used for ML models training based on different feature spaces. In general, the sub-fold fluctuation of RMSE or R² scores had a similar tendency for different feature spaces. A ML model with a smaller sub-fold fluctuation means its predictive performance is more robust. From this point, feature selection had little impact on the robustness of the ML models for $B_s$ or $D_{max}$ prediction.

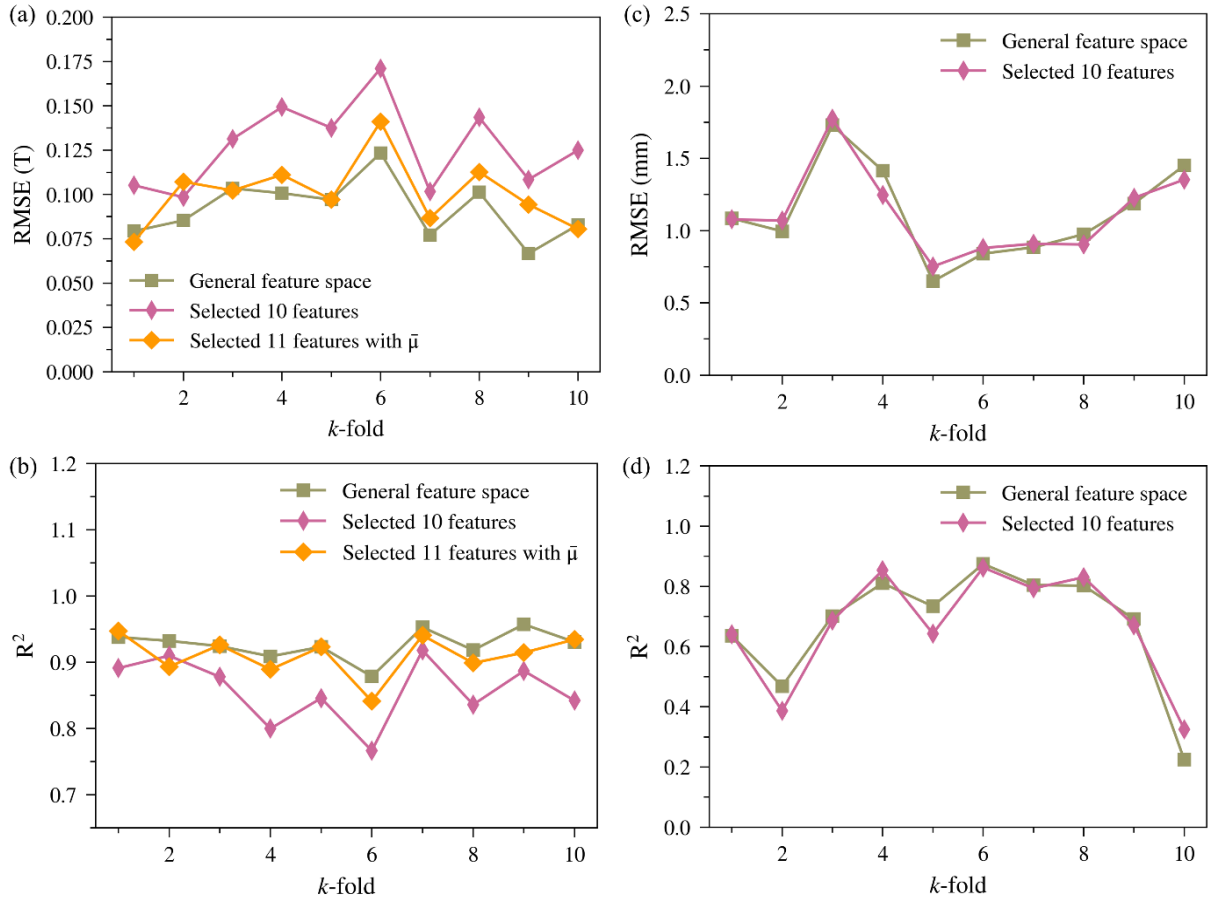

Figure 5. Detailed scores in 10-folds cross validation: (a) RMSE and (b) R² for $B_s$ prediction; (c) RMSE and (d) R² for $D_{max}$ prediction.

3.3 $D_{max}$ classification Since $D_{max}$ distribution in the dataset is too unbalanced to achieve regression analysis with high accuracy, classification analysis was conducted. A binary classification task was presented by setting the critical $D_{max}$ value to be 3 mm, which means that the whole dataset was divided into two parts, i.e., positive samples ($D_{max}$ > 3 mm) and negative samples ($D_{max}$ < 3 mm). The 5 ML algorithms used in previous regression were also suitable for classification. It should be noted that the corresponding classification algorithm of SVR is named support vector classifier (SVC) [37]. They are different implementations of support vector machines. The classification performance of different ML models was also compared by the 10-fold cross validation strategy. As shown in Figure 6(a), the classification performance of the five ML algorithms was comparable, and RF reached the highest classification accuracy of 85.9%. In detail, the cross-validated classification results of RF were expressed as a confusion matrix in Figure 6(b), where FP is false positive, TP is true positive, TN is true negative, FN is false negative. The prediction accuracy was calculated by the ratio of correctly classified samples. For example, 87.7% at the upper right corner square in Figure 6(b) means that 207 out of 252 positive samples were correctly classified by the trained RF classification model. ML models showed good classification accuracy, which could guide the discovery of new soft magnetic MGs.

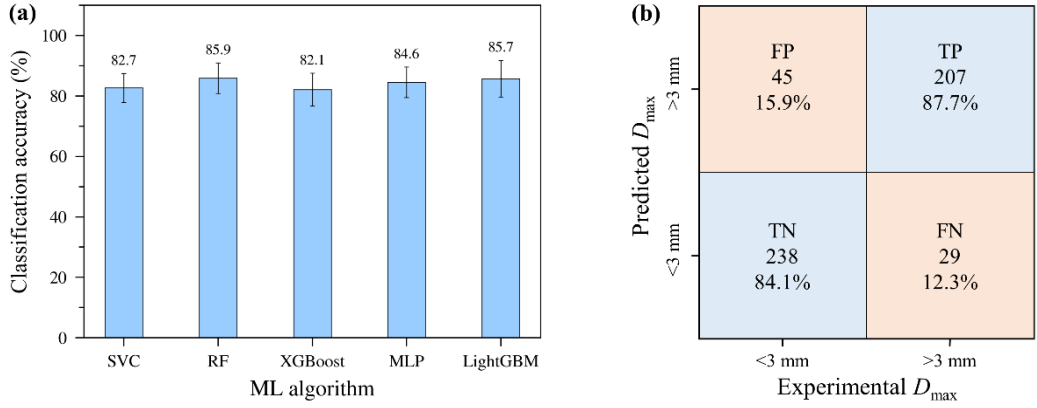

Figure 6. GFA binary classification: (a) cross-validation; (b) confusion matrix of RF.

## 4. Conclusions

(1) Based on two datasets and a general feature space, ML models can be trained to predict GFA ($D_{max}$) and magnetic property ($B_s$) of soft magnetic MGs.

(2) Among five ML algorithms, XGBoost showed the best predictive performance for both $D_{max}$ and $B_s$ prediction.

(3) Knowledge-based feature selection can greatly reduce the number of features without significant accuracy loss.

(4) With limited dataset quality, treating $D_{max}$ prediction as a classification problem was more practical than regression problem.

(5) The predictive accuracy of the trained ML models for $B_s$ and $D_{max}$ was much higher than traditional estimation methods based on physical principles.

## Acknowledgment

This work was financially supported by the Fundamental Research Funds for the Central Universities and the Beijing Advanced Innovation Center for Big Data-based Precision Medicine.